\documentclass[aps,pre,twocolumn,amsmath,amssymb,nofootinbib,superscriptaddress,showpacs,longbibliography]{revtex4-2}

\usepackage{graphicx}
\usepackage{latexsym}
\usepackage{amsmath, amssymb, amsthm}
\usepackage{mathtools}
\usepackage{xcolor}
\usepackage{comment}
\usepackage{wasysym}
\usepackage{slashed}
\usepackage{braket}

\usepackage[bottom]{footmisc}

\usepackage{microtype}
\usepackage{epigraph}
\usepackage{lipsum}

\usepackage{float}

\usepackage[normalem]{ulem}

\usepackage{appendix}

\usepackage{hyperref}

\newcommand{\kmin}{k_{\text{min}}}
\newcommand{\avg}[1]{\left\langle{#1}\right\rangle}

\begin{document}

\title{How universal is the mean-field universality class for percolation in complex networks?}

\author{Lorenzo Cirigliano}
\affiliation{Dipartimento di Fisica, Universit\`a \href{https://ror.org/02be6w209}{La Sapienza}, P.le
  A. Moro, 2, I-00185 Rome, Italy}


\date{\today}

\begin{abstract}
Clustering and degree correlations are ubiquitous in real-world complex networks. Yet, understanding their role in critical phenomena remains a challenge for theoretical studies.
Here, we provide the exact solution of site percolation in a model for strongly clustered random graphs, with many overlapping loops and heterogeneous degree distribution. We systematically compare the exact solution with heterogeneous mean-field predictions obtained from a treelike random rewiring of the network, which preserves only the degree sequence. Our results demonstrate a nontrivial interplay between degree heterogeneity, correlations and network topology, which can significantly alter both the percolation threshold and the critical exponents predicted by the heterogeneous mean-field.
These findings reveal limitations of heterogeneous mean-field theory, demonstrating that the degree distribution alone is insufficient to determine universality classes in complex networks with realistic structural features.

\end{abstract}

\maketitle

\section{Introduction}
Percolation theory provides a fundamental framework for understanding the structural properties and dynamical processes in complex networks, and it constitutes a central research topic in statistical physics and complex systems~\cite{araujo2014recent}. In its simplest form, percolation involves randomly occupying nodes (site percolation) or edges (bond percolation) with probability $\phi$, and studying the emergence of connected components as $\phi$ varies. A phase transition occurs at the percolation threshold $\phi_c$, characterized by the appearance of a giant component containing a macroscopic fraction of the network nodes. This transition has profound implications for numerous collective phenomena, including epidemic spreading~\cite{cardy1985epidemic,kenah2007second}, network robustness~\cite{callaway2000network,holme2002attack}, and cascading failures in technological infrastructures~\cite{buldyrev2010catastrophic}.

Percolation transitions are typically characterized, close to the percolation threshold, by universal scaling laws and a set of critical exponents, as it happens for standard continuous phase transitions~\cite{goldenfeld2018lectures}. The critical exponents define universality classes, groups of systems that display the same critical behaviour despite having different microscopic structures. In finite-dimensional lattices, these properties have been extensively studied \cite{stauffer2018introduction}, and percolation provides a fundamental testing ground for advanced theoretical approaches~\cite{angelini2025bethe}.
The situation is different for complex networks. Percolation critical exponents for uncorrelated random graphs with arbitrary degree distributions have been computed, revealing that the critical behaviour is strongly affected by the heterogeneity of the degree distribution~\cite{cohen2002percolation,dorogovtsev2008critical,cirigliano2024scaling}. These exponents define the heterogeneous mean-field universality class.

Uncorrelated random graphs present, however, some intrinsic limitations. Consider, for instance, the clustering coefficient~\cite{watts1998collective}, which quantifies the tendency of a node's neighbours to connect with each other. Random graphs exhibit vanishing clustering coefficient in the infinite-size limit~\cite{latora2017complex}, while real-world networks typically have high clustering. More generally, real-world complex networks exhibit heterogeneous degree distributions, community structures, degree correlations, and many short loops~\cite{latora2017complex}.

The interplay between clustering, degree correlations, and percolation has attracted considerable attention in recent years~\cite{allard2015general}. However, most prior analyses have focused primarily on the percolation threshold~\cite{serrano2006percolation,miller2009percolation,gleeson2009bond,coupette2022exactly,coupette2025universally}. While there is evidence of the key role played by degree correlations~\cite{goltsev2008percolation}, the effect of clustering on critical exponents and scaling behaviour remains less understood.

Here we investigate to what extend the heterogeneous mean-field correctly captures the percolation critical properties in complex networks. To do so, we present the exact solution of site percolation in a model for random graphs with scale-free degree distribution, many overlapping short loops, and assortative correlations. The structure of these strongly clustered networks is non-treelike, yet the solution is obtained via a mapping to another exactly solvable percolation process. We show that the degree heterogeneity is not enough to determine the critical properties, and clustering and degree correlations can alter not only the position of the percolation threshold, but also the critical exponents. Our findings 
emphasize the crucial role of the interplay between the local loop structure and the degree heterogeneity in determining the universality class for critical phenomena in complex networks.

\section{Strongly clustered random graphs via static triadic closure}
\begin{figure}
\centerline{\includegraphics[width=0.475\textwidth]{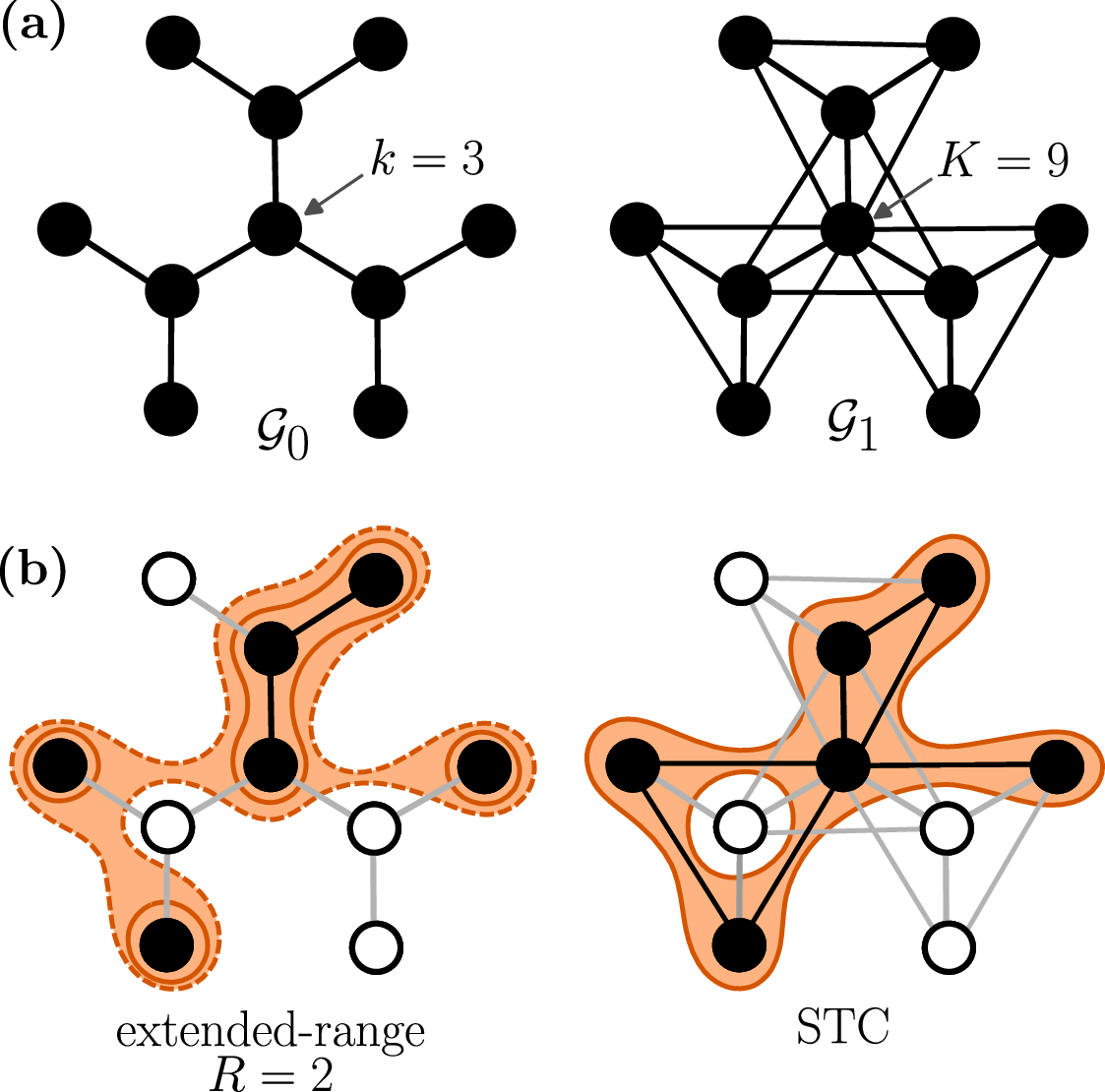}}
\caption{(a) Construction of the STC random graph via triadic closure with $f=1$, from a backbone network $\mathcal{G}_0$ (left), to the clustered graph $\mathcal{G}_1$ (right). Note that the degree of a node with original degree $k$ becomes $k+\sum_{i=1}^k r_i$ (for the node indicated by the arrow, $k=3$, $r_1=r_2=r_3=2$, and $K=9$). (b) Equivalence between extended-range percolation with $R=2$ (left) and standard percolation in the STC graph (right). Some nodes are inactive (empty circles). On the left, the dashed lines denote the extended-range connected component with $R=2$ in the backbone $\mathcal{G}_0$. The standard connected components ($R=1$) are also represented with continuous lines for comparison. On the right, continuous lines denote the standard connected component in $\mathcal{G}_1$. Note its equivalence with the $R=2$-connected component on the left.}
\label{fig.1}
\end{figure}

In this Letter, we consider a model for highly clustered networks called Static Triadic Closure (STC) random graph~\cite{cirigliano2024strongly}. The key idea behind the construction of these random graphs is simple: take a backbone graph $\mathcal{G}_0$, and close each triad with probability $f$, to create a new random graph $\mathcal{G}_f$. Note that $\mathcal{G}_f$ has an intricate local structure, with loops of arbitrary length and overlapping motifs, even if the backbone is a perfect tree. These features mark an important distinction between STC graphs and most of the exactly solvable models for clustered networks in the literature. Such models typically rely on some sort of treelike structure -- such as trees of loops~\cite{newman2009random}, trees of cliques~\cite{gleeson2009bond} or trees of partial cliques~\cite{karrer2010random} -- while overlapping loops of arbitrary length are usually not included, as they constitute theoretical difficulties which are hard to overcome by means of message passing techniques~\cite{cantwell2019message}. STC networks present instead all these features. Yet, they can be exactly solved if the backbone $\mathcal{G}_0$ is treelike: in this case, even if message passing dramatically fails on $\mathcal{G}_f$, it works on the backbone.
The analysis of the topological properties (clustering coefficient, motifs densities) of STC random graphs for arbitrary $f$, obtained from uncorrelated random backbones with arbitrary degree distribution $p_k$, has been developed in~\cite{cirigliano2024strongly}, where it has been shown that these graphs may have strong clustering and an intricate non-treelike structure in the infinite-size limit. They also exhibit assortative correlations, meaning that nodes of similar degrees are more likely to be connected within each others. Here we focus only on the case $f=1$, \textit{i.e.}, on random graphs $\mathcal{G}_1$ obtained closing all the triads in the backbone, see Fig.~\ref{fig.1}(a). Furthermore, an important ingredient that we want to include in our analysis is the degree heterogeneity. For this reason, we consider the case of uncorrelated random backbones with a power-law degree distribution $p_k \sim k^{-\gamma_d}$ for large $k$. The resulting STC random graph $\mathcal{G}_1$ has, in the infinite-size limit, a degree distribution $P_K \sim K^{-\widetilde{\gamma}_d}$ for large $K$, with $\widetilde{\gamma}_d = \gamma_d-1$. This can be easily seen as follows. A node with original degree $k$ surrounded by neighbours with original residual degrees $r_1,\dots, r_k$, has exactly degree $K=k+\sum_{i=1}^{k}r_i$. The residual degrees are uncorrelated random variables distributed according to $q_r=(r+1)p_{r+1}/\langle k \rangle$. The conditional degree distribution is then simply $P_{K|k,r_1,\dots,r_k}=\delta_{K,k+\sum_{i=1}^{k}r_i}$. Averaging over $P(k,r_1,\dots,r_k)=p_k q_{r_1}\dots q_{r_k}$, since $\mathcal{G}_0$ is uncorrelated, we then get
\begin{equation}
\label{eq:degree_distribution}
 P_K = \sum_{k, r_1,\dots,r_k}p_k q_{r_1}\dots q_{r_k} \delta_{K,k+\sum_{i=1}^{k}r_i}.
\end{equation}
Even if the r.h.s. of Eq.~\eqref{eq:degree_distribution} cannot be written explicitly, we can easily find an expression for the generating function $G_0(z)=\sum_{K}P_K z^{K}$ of the degree distribution $P_K$. Multiplying by $z^K$ and summing over $K$ we get
\begin{equation}
\label{eq:G_0_STC}
 G_0(z)=g_0\left(z g_1(z) \right),
\end{equation}
where $g_0(z)=\sum_{k}p_k z^k$ and $g_1(z)=\sum_{r}q_r z^r$. For power-law networks with degree exponent $\gamma_d$, the generating functions have a singularity in $z=1$ given, at leading order in $\varepsilon=1-z$, by $g_0(1-\varepsilon) \sim 1 - c_1\varepsilon^{\gamma_d-1}$ and $ g_1(1-\varepsilon) \sim 1 - c_2\varepsilon^{\gamma_d-2}$,
where $c_1, c_2$ are constants depending on $\gamma_d$ and in general on the low-degree part of the $p_k$. With these expansions from Eq.~\eqref{eq:G_0_STC} we get at leading order for the singular part of $G_0$
\begin{equation}
 G_0(1-\varepsilon) \sim 1 - C_1 \varepsilon^{\gamma_d-2} = 1 - C_1 \varepsilon^{\widetilde{\gamma}_d-1},
\end{equation}
where $\widetilde{\gamma}_d=\gamma_d-1$. Hence we can conclude, using asymptotic theorems for generating functions~\cite{flajolet1990singularity}, that $P_K$ has a power-law tail, asymptotically for large $K$, with degree exponent $\widetilde{\gamma}_d$. In other words, the STC mechanism decreases by $1$ the degree exponent. This is reasonable, since we are changing the degree of each node with the sum of its neighbours degrees, which are random variables distributed with probability proportional to $k p_k$.

\section{Extended-range percolation with $R=2$ and STC graphs}
Before getting to the solution of percolation in STC graphs, we need to briefly discuss another percolation model whose exact solution is known. Consider a network in which nodes are active with probability $\phi$. The standard notion of connectivity between two active nodes is equivalent to the existence of a path made of consecutive active nodes between them. In extended-range percolation~\cite{cirigliano2023extended,cirigliano2024general} a range $R$ is introduced, and each active node is connected to all other active nodes at topological distance at most $R$. With this definition, inactive nodes are allowed along the paths connecting active nodes. In particular, all paths containing at most $R-1$ consecutive inactive nodes can be used to connect active nodes. See Fig.~\ref{fig.1}(b) for a pictorial representation for $R=2$.

The reader may now wonder why we are introducing another percolation model. The reason is the following. Extended-range percolation with $R=2$ in a graph $\mathcal{G}_0$ is exactly equivalent to standard site percolation in the graph obtained from $\mathcal{G}_0$ by closing all its open triads, which is precisely the STC graph $\mathcal{G}_1$ presented in the previous section, see Fig.~\ref{fig.1}(b). Hence, the exact solution of extended-range percolation in $\mathcal{G}_0$ is also the exact solution of standard site percolation in the clustered and correlated network $\mathcal{G}_1$. In the rest of this section we derive the exact solution of extended-range percolation with $R=2$ for uncorrelated locally treelike graphs $\mathcal{G}_0$ with a given $p_k$.

Equations for the order parameter $P^{\infty}$ for extended-range percolation have already been derived~\cite{cirigliano2023extended,cirigliano2024general}. In particular, the formalism presented in \cite{cirigliano2024general} is general, as it solves the problem for any $R$. However, for the sake of simplicity, we follow here the solution discussed in~\cite{cirigliano2023extended}.
The key idea in~\cite{cirigliano2023extended} to solve the problem for $R=2$ is to introduce two probabilities $u_1$ and $u_2$, representing the probabilities of not reaching the giant component following a random edge which ends in an active or inactive node, respectively. In standard percolation, $u_2=1$. In extended-range percolation instead, there is a nonvanishing probability of reaching the GC if we reach an inactive node. The important thing is to not get too many -- in particular, $2$ for $R=2$ -- consecutive inactive nodes.

Here, we focus on the cluster size distribution $\pi(s)$, the probability that an active node belongs to a small component of size $s$. In particular, we solve for its generating function $H_0(z)=\sum_{s} \pi(s) z^s$. The knowledge of $H_0(z)$ provides us, after some cumbersome but straightforward computations, a complete characterisation of the critical properties~\cite{cirigliano2024scaling}.
In analogy with the probabilities $u_1$ and $u_2$, we define $\rho_1(s)$ and $\rho_2(s)$ as the probabilities that following a randomly chosen link leading to an active or inactive node, respectively, we reach a small cluster of size $s$. With these two quantities we can write an equation for $\pi(s|k)$, taking a node of degree $k$ and then averaging over $p_k$ to get $\pi(s)=\avg{\pi(s|k)}$ as
\begin{equation}
\label{eq:pi_s_STC}
 \pi(s) = \avg{ \sum_{\{\xi_1, \dots, \xi_k \}|\mathcal{C}_1 } \sum_{n=0}^{k} \mathcal{B}_{n}^{k}(\phi) \prod_{i=1}^{n} \rho_{1}(\xi_i) \!\! \prod_{j=n+1}^{k} \!\! \rho_2(\xi_j) },
\end{equation}
where $\mathcal{B}_{n}^{k}(\phi)={k \choose n} \phi^n(1-\phi)^{k-n}$ is the probability of a configuration of the percolation process with $n$ active neighbours among the $k$ available, each active branch $i$ taking a factor $\rho_1(\xi_i)$ and each inactive branch $j$ taking a factor $\rho_2(\xi_j)$, summed over all possible values of $n$, and the sum over the variables $\xi_1, \dots, \xi_k$ has a constraint $\mathcal{C}_1$: $\sum_{i=1}^{k}\xi_i = s-1$, in order to ensure that the size of the small component of the active node we picked at random is $s$. Again, Eq.~\eqref{eq:pi_s_STC} cannot be solved explicitly, but we can find an expression for the generating function $H_0(z)$, multiplying by $z^s$ and summing over $s$. The result is, after a tedious but straightforward computation\footnote{Write the constraint over the $\xi_i$ variables as a Kronecker delta $\delta_{s-1, \sum_{i=1}^{k}\xi_i}$; sum over $s$ first and distribute the factors $z^{\xi_i}$ one for each branch; sum over the variables $\xi_i$, using the independence of the branches, and use the definition of $H_1(z)$ and $H_2(z)$; sum over $n$ using the binomial formula; finally, average over $p_k$.}
\begin{equation}
\label{eq:H_0_STC}
H_0(z)=z g_0\left(\phi H_1(z)+(1-\phi)H_2(z) \right),
\end{equation}
where we introduced the generating functions $H_1(z)=\sum_{s}\rho_1(s)z^s$ and $H_2(z)=\sum_{s}\rho_2(s)z^s$.
The same argument applies for the distribution $\rho_1(s)$. If we condition on active nodes of excess degrees $r$, we can write $\rho_1(s)=\avg{\rho_1(s|r)}$ with the average performed over $q_r$. Hence $\rho_1(s)$ is given by Eq.~\eqref{eq:pi_s_STC} with $k$ replaced by $r$ and the average taken over $q_r$. This gives us for the generating function
\begin{equation}
\label{eq:H_1_STC}
 H_1(z) = z g_1\left(\phi H_1(z)+(1-\phi)H_2(z) \right).
\end{equation}
A bit more care must be taken for $\rho_2(s)$. Since $R=2$, and two consecutive inactive nodes are not allowed, when writing the analogous of Eq.~\ref{eq:pi_s_STC} for $\rho_2(s)$ we must consider that if another inactive node is encountered, that certainly leads to a component of size $0$. This means to substitute the factors $\rho_2(s_i)$ associated to each inactive branch with $\delta_{s_i,0}$.
Also, the constraint, denoted with $\mathcal{C}_2$, on the variables $\xi_1, \dots, \xi_r$ is that they sum to $s$, not to $s-1$, since we don't want to count inactive nodes as part of small components. This gives us
\begin{equation}
 \rho_2(s)=\avg{ \sum_{\{\xi_1, \dots, \xi_r \}|\mathcal{C}_2 } \sum_{n=0}^{r} \mathcal{B}_{n}^{k}(\phi)  \prod_{i=1}^{n} \rho_{1}(s_i) \prod_{j=n+1}^{r} \delta_{s_j,0} },
\end{equation}
again the average being performed over $q_r$, from which we get
\begin{equation}
\label{eq:H_2_STC}
 H_2(z) = g_1\left(\phi H_1(z) + 1- \phi \right).
\end{equation}
Now we have an implicit expression for the generating function of the cluster size distribution $H_0(z)$ via Eqs.~\eqref{eq:H_0_STC},~\eqref{eq:H_1_STC}, and \eqref{eq:H_2_STC}. We can conclude our digression on extended-range percolation and come back to the percolation in the clustered STC networks.

\section{Percolation in STC random graphs}

In this section, we take advantage of the exact mapping between extended-range percolation with $R=2$ in $\mathcal{G}_0$ and standard site percolation in $\mathcal{G}_1$. 
In particular, we can use Eqs.~\eqref{eq:H_0_STC},\eqref{eq:H_1_STC}, and \eqref{eq:H_2_STC} to compute various percolation observables in $\mathcal{G}_1$. Working out their behaviour close to the percolation threshold, we get exact expressions for the critical exponents. An important remark is in order here. The equations for extended-range percolation are expressed in terms of $\gamma_d$, the exponent of the backbone degree distribution $p_k$, while we need instead to express our results in terms of $\widetilde{\gamma}_d$, the exponent of the degree distribution of the STC networks. This is simply achieved by setting $\gamma_d=\widetilde{\gamma}_d+1$.

\begin{figure}
\centerline{\includegraphics[width=0.475\textwidth]{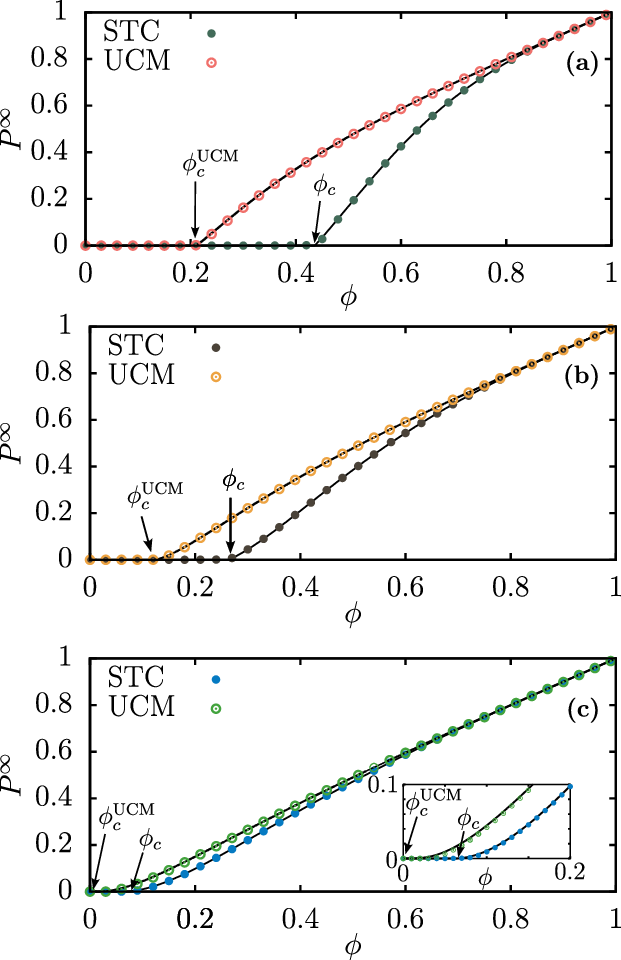}}
\caption{Numerical simulations of site percolation in STC random graphs (full symbols) and in a random rewiring of these networks (UCM) (empty symbols), keeping the same degree sequence but with no correlations nor the short-loop structure. The synthetic networks considered have $N=10^6$ nodes, $\kmin=2$ and (a) $\widetilde{\gamma}_d=4.5$, (b) $\widetilde{\gamma}_d=3.5$, (c) $\widetilde{\gamma}_d=2.5$. Results are averaged over $100$ independent realizations. Continuous lines are the exact solution for site percolation in STC, Eqs.~\eqref{eq:order_parameter}, and for site percolation in the UCM counterpart, Eq.~\eqref{eq:order_parameter_UCM}. }
\label{fig.2}
\end{figure}

We begin considering the order parameter $P^{\infty}$, which behaves as $P^{\infty}/\phi \sim (\phi-\phi_c)^{\beta}$ close to criticality. Since active nodes are either in the giant component (GC) or in small components of any size, we have $\phi = \phi H_0(1)+P^{\infty}$. Setting $u_1=H_1(1)$ and $u_2=H_2(1)$,
\begin{equation}
\label{eq:order_parameter}
 P^{\infty}=\phi \left[1-g_0\left(\phi u_1 +(1-\phi)u_2 \right) \right],
\end{equation}
where $u_1$ and $u_2$ are solution of the recursive equations
\begin{align}
\label{eq:u_1}
 u_1 &= g_1(\phi u_1 + (1-\phi) u_2),\\
 \label{eq:u_2}
 u_2 &= g_1(\phi u_1 + 1-\phi),
\end{align}
thus recovering the equations in~\cite{cirigliano2023extended}. Numerical solutions of Eq.~\eqref{eq:order_parameter}-\eqref{eq:u_2} are presented in Fig.~\ref{fig.2} for various values of the degree exponent $\widetilde{\gamma}_d$, compared with results from numerical simulations on STC graphs. The trivial solution $u_1=u_2=1$, corresponds to $P^{\infty}=0$. Such a solution is stable until the largest eigenvalue $\Lambda$ of the Jacobian of Eqs.~\eqref{eq:u_1},\eqref{eq:u_2}, evaluated in $u_1=u_2=1$, is smaller than $1$. The criticality condition is then given by $\Lambda(\phi_c)=1$, from which we get $\phi_c = (1+b - \sqrt{(1+b)^2-4})/(2b)$, where $b=g_1'(1)=\langle k(k-1)\rangle/\langle k \rangle$. Remarkably, this expression coincides with the one found in~\cite{eggarter1975percolation} for the case of Cailey-tree backbones with degree $b+1$. Using $\avg{K}=G_0'(1)=\avg{k}+\avg{k(k-1)}=\avg{k}(1+b)$, we can write the percolation threshold as
\begin{equation}
\label{eq:threshold}
 \phi_c = \frac{\avg{K}-\sqrt{\avg{K}^2-4\avg{k}^2}}{2\left(\avg{K}-\avg{k}\right)},
\end{equation}
where $\avg{k}$ is the average degree of the backbone $\mathcal{G}_0$ and $\avg{K}$ is the average degree of $\mathcal{G}_1$. From this expression, we see that the threshold is always positive if $\avg{K}< \infty$, even for $2 < \widetilde{\gamma}_d <3$, when $\langle K^2 \rangle$ diverges. The analysis of Eqs.~\eqref{eq:order_parameter}-\eqref{eq:u_2} close to the percolation threshold gives us the values of the critical exponent $\beta$ as a function of $\widetilde{\gamma}_d$. Details of this computation are reported in the SM, and the results are presented in Table~\ref{tab.1}.

We turn now our attention to the average size of small components, defined as $\avg{s}=\sum_{s}s \pi(s) / \sum_{s'}\pi(s')$. This quantity scales, for $\phi$ close to the percolation threshold, as $\avg{s}-1 \sim |\phi - \phi_c|^{-\gamma}$. Using Eq.~\eqref{eq:H_0_STC} we can write
\begin{equation}
\langle s \rangle = \frac{\sum_s s \pi(s)}{\sum_s' \pi(s')}  = \frac{H_0'(1)}{H_0(1)}. 
\label{eq:average_s}
\end{equation}

A straightforward computation, whose details can be found in the Supplementary Material (SM), allows us to get an exact expression for $\langle s \rangle$, whose behaviour for $\phi$ close to $\phi_c$ gives us the critical exponent $\gamma$. In Table~\ref{tab.1} we report the values of the exponent $\gamma$ for different degree exponents $\widetilde{\gamma}_d$.

Finally, we consider the cutoff size $s_{\xi}$ and the critical exponent $\sigma$. Formally, $s_{\xi}$ is defined from the scaling behaviour of the cluster size distribution for $\phi$ close to the percolation threshold, as $\pi(s) \sim s^{-\tau+1}\exp(-s/s_{\xi})$, where $s_{\xi} \sim |\phi - \phi_c|^{-1/\sigma}$, and $\tau$ is another critical exponent. Using the scaling relation $\sigma=1/(\beta+\gamma)$ we get the results reported in Table~\ref{tab.1}. As a consistency check, we also computed the exponent $\sigma$ independently from the others. In fact, the cutoff size $s_{\xi}$ can be explicitly computed from the knowledge of $H_0(z)$, as it is related to the position of the closest singularity to the origin in the complex plane. See the SM for details.


\begin{widetext}
\begin{table}
\caption{Values of the critical exponents for site percolation on STC random graphs, and comparison with the predictions of the heterogeneous mean field (denoted with the subscript hMF). In the last row networks are no longer sparse and hMF is not applicable.}
    \label{tab.1}
    \begin{tabular}{|c|c|c|c|c|c|c|c|c|} \hline
      degree exponent & $\phi_c$ & $\beta$  &  $\gamma$ & $\sigma$  &$\phi_c^{\text{UCM}}$ & $\beta_{\text{hMF}}$  &  $\gamma_{\text{hMF}}$ & $\sigma_{\text{hMF}}$  \\ \hline
      $\widetilde{\gamma}_d > 4 $ & $>0$ & $1$  & $1$ & $\frac{1}{2}$ & $>0$ &  $1$ & $1$ &  $\frac{1}{2}$  \\
      $3<\widetilde{\gamma}_d < 4$ & $>0$ &  $1$  & $1$ & $\frac{1}{2}$ & $>0$ & $\frac{1}{\widetilde{\gamma}_d-3}$ & $1$ &  $\frac{\widetilde{\gamma}_d-3}{\widetilde{\gamma}_d-2}$  \\
      $2< \widetilde{\gamma}_d <3$  & $>0$ & $\frac{1}{\widetilde{\gamma}_d-2}$ & $1$ & $\frac{\widetilde{\gamma}_d-2}{\widetilde{\gamma}_d-1}$ & $=0$ & $\frac{1}{3-\widetilde{\gamma}_d}$ & $-1$ &   $\frac{3-\widetilde{\gamma}_d}{\widetilde{\gamma}_d-2}$ \\
      $1<\widetilde{\gamma}_d<2$ & $=0$ &  $\frac{\widetilde{\gamma}_d-1}{1-(\widetilde{\gamma}_d-1)^2}$ & $-\frac{(2-\widetilde{\gamma}_d)(\widetilde{\gamma}_d-1)}{1-(\widetilde{\gamma}_d-1)^2}$ & $\frac{1-(\widetilde{\gamma}_d-1)^2}{(\widetilde{\gamma}_d-1)^2}$ & N.A. &  N.A. & N.A. &  N.A.  \\ \hline
    \end{tabular}
\end{table}
\end{widetext}

\section{Comparison with the heterogeneous mean-field}
The purpose of this section is to see to what extend degree heterogeneity, clustering and degree correlations affect the critical exponents. In order to get some insight, we compare the exact solution for site percolation in STC graphs derived above with the heterogeneous mean-field predictions, for $\widetilde{\gamma}_d>2$, which ignores clustering and degree correlations. To do so, we perform a random rewiring of the clustered network $\mathcal{G}_1$. This procedure preserves the degree sequence and destroys the short loop structure, producing an uncorrelated locally treelike maximally random graph whose degrees are distributed according to $P_K$. In other words, if the network $\mathcal{G}_1$ is sparse hence $\widetilde{\gamma}_d>2$, we create locally treelike random graphs with the same degree sequence of the STC graph. 
For such sparse networks, the exact solution of site percolation is known~\cite{callaway2000network}, and we simply have
\begin{equation}
\label{eq:order_parameter_UCM}
P^{\infty}_{\text{UCM}}=\phi \left[1-G_0\left(1-\phi + \phi u \right) \right],
\end{equation}
where $u$, the probability of not reaching the GC following a random link ending in an active node, is the stable solution of the fixed-point equation
\begin{equation}
\label{eq:u_UCM}
 u = G_1(1-\phi+\phi u),
\end{equation}
where $G_1(z)=G_0'(z)/G_1'(1)$. A similar formula exists for the average cluster size $\avg{s}_{\text{UCM}}$. See Fig.~\ref{fig.2} for a comparison between Eq.~\eqref{eq:order_parameter_UCM}, the exact solution of percolation in the clustered STC graphs Eq.~\eqref{eq:order_parameter}, and numerical simulations on large networks and various values of the degree exponent $\widetilde{\gamma}_d$.

The percolation threshold is given by $\phi_c^{\text{UCM}}=1/B$, where $B=G'_1(1)=\avg{K(K-1)}/\avg{K}$ is $\mathcal{G}_1$'s branching factor. Note that from Eq.~\eqref{eq:G_0_STC} we can compute the moments of $P_K$, as $\avg{K(K-1)}=G_0''(1)$ and $\avg{K}=G_0'(1)$, hence we have an explict expression of the threshold $\phi_c^{\text{UCM}}$ in terms of the moments of the backbone degree distribution $p_k$. Since $\avg{(K(K-1))}$ depends on $\avg{k(k-1)(k-2)} \sim \avg{k^3}$, the branching factor $B$ diverges for $\gamma_d<4$, hence for $\widetilde{\gamma}_d<3$, and the threshold vanishes in the infinite-size limit, as expected for scale-free uncorrelated random graphs. Percolation critical exponents are known in this case~\cite{cohen2002percolation,dorogovtsev2008critical,cirigliano2024scaling}, and they are reported in Table~\ref{tab.1}. It is also well-known that for $2<\widetilde{\gamma}_d<4$ the singularity in the generating function of power-law distributions is strong enough to interfere with the critical behavior, leading to $\widetilde{\gamma}_d$-dependent critical exponents. This defines the so-called heterogeneous mean-field (hMF) universality class~\cite{cohen2002percolation,dorogovtsev2008critical,cirigliano2024scaling}.

Let us now analyze the results in Table~\ref{tab.1}. For $\widetilde{\gamma}_d>4$, the only difference between the exact solution of percolation in STC networks and the heterogeneous mean-field concerns the prediction of the threshold, since $\phi_c > \phi_{c}^{\text{UCM}}$. Degree correlations and clustering increase the threshold with respect to the UCM, while critical exponents are unchanged. However, the effect of assortative correlations is to lower the percolation threshold~\cite{goltsev2008percolation}. Hence, we can conclude that the presence of loops is actually the crucial ingredient here, not the assortativity. Note that these results are also valid for STC networks with homogeneous degree distributions, which can be created using a homogeneous backbone, such as random regular graphs or Erd\H{o}s-R\'enyi graphs. Even if the local loopy structure of the STC graph makes it completely different from the treelike configuration model, causing the shift in the percolation threshold, the critical exponents are the homogeneous mean-field ones. We can conclude that in this case the mean-field universality class is robust under variations of the loop and correlation structure, even if the locally treelike property is completely lost. Increasing the degree of heterogeneity in the STC networks, for $3<\widetilde{\gamma}_d<4$, we still have $\phi_c > \phi_{\text{UCM}}$. However, in this regime hMF predicts $\widetilde{\gamma}_d$-dependent critical exponents, while the critical exponents in the STC networks remain the homogeneous mean-field ones. It is clear that this effect is ultimately due to the existence of an underlying locally treelike backbone with exponent $\gamma_d=\widetilde{\gamma}_d+1>4$, whose critical properties are indeed described by the homogeneous mean-field critical exponents. Yet, it is remarkable and surprising that the critical properties of an intricate structure such as the STC one are captured by the backbone alone. The role of clustering and degree correlations is even stronger when degree heterogeneity is stronger, for $2<\widetilde{\gamma}_d<3$. Here, not only the critical exponents are different from the hMF, but we also have $\phi_c>0$, while the threshold vanishes in the UCM, since $B \sim \langle K^2 \rangle$ diverges. Ultimately, we notice that our solution is valid even for $1<\widetilde{\gamma}_d<2$, when the networks are no longer sparse, as the average degree $\langle K \rangle$ diverges in the infinite-size limit. Remarkably, even if heterogeneous mean-field does not apply here, we are providing the exact solution of percolation on a non-sparse network.

\section{Discussion}

\begin{figure}
\centerline{\includegraphics[width=0.475\textwidth]{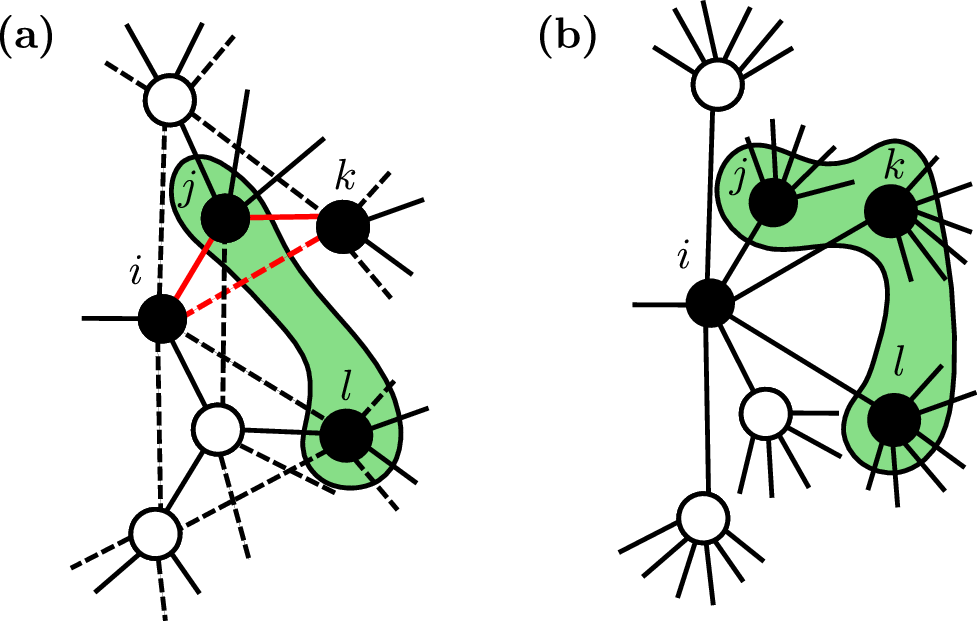}}
\caption{(a) Modified branching process for percolation in STC graphs. Here, the edges of the backbone are represented by solid lines, and the edges created by the STC mechanism are represented by dashed lines. The first generation of offsprings emanating from $i$ (green shaded region) is formed by the active nodes that are $i$'s nearest neighbors in $\mathcal{G}_0$ (node $j$), and the active nodes that are $i$'s second neighbors in $\mathcal{G}_0$ and are reachable in $\mathcal{G}_1$ only by dashed lines (node $l$). Note that this modified branching process avoids overcounting effects caused by the presence of loops: node $k$, which is a common neighbor of both $i$ and $j$ (red triangle) belongs to the second generation of offsprings. (b) The standard branching process approach in a random rewiring of the same network, in which the loop structure has been destroyed. The first generation of offsprings emanating from node $i$ are nodes $j,k,l$. This difference with the modified branching process in panel (a) is crucial.}
\label{fig.3}
\end{figure}

The comparison between the exact solution and the heterogeneous mean-field in Table~\ref{tab.1} reveals that that site percolation on the STC networks with degree exponent $\widetilde{\gamma}_d$ presents deviations from the heterogeneous mean-field theory. This suggests that the interplay between the local loopy structure and the degree heterogeneity is not well-captured by the hMF alone. To understand the physical mechanisms behing this deviation from the hMF, it is instructive to consider the percolation process on the STC graphs from a branching process perspective~\cite{coupette2022exactly,coupette2025universally}. While a standard branching process approach fails on $\mathcal{G}_1$ due to its many loops and complex structure, a modified branching process that accounts for the network's construction can be defined as follows. With the help of Fig.~\ref{fig.3}(a), let us consider the number of offsprings, avoiding potential overcounting due to the loops, emanating from an active node $i$. This is given by the number of active excess neighbors of node $i$ in $\mathcal{G}_0$, plus the active excess neighbors of $i$'s inactive excess neighbors in $\mathcal{G}_0$. Note that in this way we are correctly taking into account for the presence of loops. The generating function for this modified branching process is then given by
\begin{equation}
 \widetilde{G}_1(z)=g_1\left(\phi z + (1-\phi) g_1(1-\phi + \phi z) \right).
\end{equation}
The quantity $\widetilde{B}(\phi)=\widetilde{G}_1'(1)$ is the average number of offsprings without overcounting. The criticality condition of branching processes $\widetilde{B}(\phi_c)=1$ allows us to recover Eq.~\eqref{eq:threshold}. The fixed point equation $u=\widetilde{G}_1(u)$ gives us the probability of branching in finite clusters, and it is the same as Eq.~\eqref{eq:u_1} once we replace $u_2$ with its expression in Eq.~\eqref{eq:u_2}. Following a similar argument, it is possible to derive equations for $H_0(z)$ and $H_1(z)$, and to compute the various percolation observables. Despite the perfect equivalence with the formalism used for extended-range percolation, this interpretation in terms of branching processes provides a clear physical picture. If we ignore the presence of loops using heterogeneous mean-field as in Fig.~\ref{fig.3}(b), we overcount the actual number of offsprings. This leads to the wrong prediction of the percolation threshold, since $G_1'(1)=\phi B$ predicted by the hMF is much larger than the effective average number of offsprings $\widetilde{B}(\phi)$, even for bounded degree distributions. However, when degree heterogeneity is present, and in particular when the number of second neighbors in $\mathcal{G}_0$ has unbounded fluctuations, the error we commit by neglecting the loops can affect also the critical properties. Thus, while loops alone are not enough to observe deviations from hMF in the critical exponents for percolation in STC networks, the combined effect of strong degree heterogeneity and short loops leads to a different critical behavior from the predictions of the hMF theory.

In conclusion, in this Letter we have presented the exact solution of site percolation in a model of strongly clustered random graphs with heterogeneous degree distribution and assortative degree correlations.
The key feature of the model analysed in this Letter, which makes it different from most of the models for clustered networks in the literature, is the presence of many overlapping loops -- loops sharing common edges. By comparing percolation in these clustered networks and in random rewired networks with the same degree sequence -- destroying clustering and degree correlations -- we have shown that that the degree exponent $\widetilde{\gamma}_d$ alone is not enough to determine the universality class. Our results reveal a nontrivial interplay between the degree exponent and the network local structure, which may affect both the percolation threshold and the critical exponents. This constitutes an example of a sparse network with assortative correlations, strong clustering and scale-free degree distribution, where the critical behaviour differs from the heterogeneous mean-field one~\cite{goltsev2003critical}.
We conclude with a conceptual remark. The key ingredient for the exact solvability of percolation in such non-treelike structures as the STC graphs is the existence of an underlying treelike backbone, which is known by construction. Furthermore, we have shown that this backbone is ultimately responsible of the critical properties of the loopy STC graph. This calls for an interesting theoretical question. As STC networks qualitatively capture well the structure of many real-world networks~\cite{cirigliano2024strongly}, we may wonder if it is possible to detect a treelike backbone of real-world networks, and use this backbone to correctly capture the critical properties of the entire network. The exploration of such an inference problem deserves further studies.

\acknowledgments
The author is grateful to C. Castellano for his precious comments on the manuscript, and to G. Tim\'ar for stimulating discussions on the subject. The author would also like to thank two anonymous referees whose useful comments have substantially improved the quality of the manuscript.

\bibliography{1_references}

\end{document}


\title{Supplementary Material for ``How universal is the mean-field universality class for percolation in complex networks?''}

\author{Lorenzo Cirigliano}
\affiliation{Department of Physics, \href{https://ror.org/02be6w209}{Sapienza University of Rome}, I-00185 Rome, Italy
}




\date{\today}


\maketitle

In this Supplementary Material, we briefly sketch a general formalism~\cite{cirigliano2024scaling} to study the percolation critical behaviour from the implicit knowledge of the generating function of the cluster size distribution. Then, we apply this formalism to the equations derived in the main text for percolation in strongly clustered random graphs. These equations -- since they are obtained via a mapping to extended-range percolation on the backbone -- are expressed in terms of the degree exponent of the backbone $\gamma_d$. The connection between the exponent of the backbone and the exponent of the STC graphs is given by $\gamma_d=\widetilde{\gamma}_d+1$. This allows us to compute all the percolation critical exponents in the strongly clustered STC graphs as function of the degree heterogeneity $\widetilde{\gamma}_d$.

\section{General formalism for percolation critical behaviour}
In the main text, we have derived recursive equations for the generating function of the cluster size distribution $H_0(z)=\sum_{s}\pi(s)z^s$. We report these equations here for completeness, as they are the starting point of our analysis. We have
\begin{align}
 H_0(z)=zg_0\big(\phi H_1(z)+(1-\phi)H_2(z)\big),
 \label{eq:H_0}
\end{align}
where $H_1(z)$ and $H_2(z)$ satisfy the following recursive equations
\begin{align}
\label{eq:H_1}
 H_1(z)&=zg_1\big(\phi H_1(z)+(1-\phi)H_2(z)\big),\\
 \label{eq:H_2}
 H_2(z)&=g_1\big(\phi H_1(z)+1-\phi\big).
\end{align}

As shown in \cite{cirigliano2024scaling}, the knowledge of $H_0(z)$, even if via an implicit equation such as Eq.~\eqref{eq:H_0}, is sufficient to solve for most of the percolation critical properties. For instance, the order parameter $P^{\infty}$ can be computed as
\begin{equation}
  P^{\infty}(\phi)=\phi \big[1-H_0(1)\big],
  \label{eqP}
\end{equation}
and the average cluster size is given by
\begin{equation}
  \langle s \rangle =
  \frac{\sum_{s} s\pi(s)}{\sum_{s'} \pi(s')}
  = \frac{H'_0(1)}{H_0(1)}
  \label{s}.
\end{equation}
Hence the quantity $H_0(z)$ for $\phi \to \phi_c$ captures the critical behaviour we are interested in as
\begin{align}
\label{eq:scaling_exponent_2}
1-H_0(1)|_{t} &\simeq B t ^{\beta},\\
\label{eq:scaling_exponent_3}
\frac{H_0'(1)|_{t}}{H_0(1)|_{t}}-1 & \simeq C_{\pm}|t|^{-\gamma},
\end{align}
where $|_{t}$ indicates that the functions are evaluated at $\phi=\phi_c+t$. The other critical exponents $\alpha$ and $\delta$ can be derived in a similar manner from $H_0(z)$, see~\cite{cirigliano2024scaling}.

A key quantity for understanding the percolation critical behaviour is the tail of the cluster size distribution~\cite{cirigliano2024scaling}
\begin{equation}
\pi(s)|_{t} \simeq q_0 s^{1-\tau}f_{\pm}(q_1 t s^{\sigma}),
\label{eq:scaling}
\end{equation}
for $s \gg 1$ and $|t| \ll 1$, where $q_0, q_1$ are nonuniversal positive
constants, $f_{\pm}(x)$ are universal scaling functions, while
$\tau$ and $\sigma$ are universal critical exponents.
The scaling functions decay to zero quickly for large values of their argument:
$f_{\pm}(x) \ll 1$ for $|x| \gg 1$.
This naturally introduces the quantity
\begin{equation}
 s_{\xi}\simeq \Sigma |t|^{-1/\sigma},
 \label{eq:definition_correlation_size}
\end{equation}
which plays the role of a correlation size, where $\Sigma=q_1^{-1/\sigma}$.
For $s/s_\xi \ll 1$, the cluster size distribution scales as
$\pi(s)_{|t}\sim s^{1-\tau}$.
This implies that, if $\sigma >0$ so that
$s_\xi \to \infty$ as $t \to 0$, at criticality $\pi(s)$ exhibits a pure power-law tail for large $s$
with exponent $1-\tau$.

From Eq.~\eqref{eq:H_0} one can also work out explictly the exponents $\tau$ and $\sigma$. Here we focus on $\sigma$, and we refer to~\cite{cirigliano2024scaling} for details about the exponent $\tau$. The convergence radius of the power series $H_0(z)=\sum_{s}\pi(s)z^s$, which is defined as the closest singularity to the origin $z^*$, is given by $z^{*}=\lim_{s \to \infty}|\left[\pi(s) \right]^{1/s} |$. Assuming an exponential decay of the form $e^{-s/s_{\xi}}$, as it always happens, we simply have
\begin{equation}
 s_{\xi}=\frac{1}{\log(z^*)}.
\end{equation}
At criticality, $\pi(s)$ exhibits a pure power-law decay, which implies $z^{*}=1$. Then from the behaviour of $z^{*}(t)$ as $t \to 0$ we can get the critical behaviour of $s_{\xi}$, \textit{i.e.}, the critical exponent $\sigma$.

We have briefly sketched the formalism needed to extract the percolation critical exponents from $H_0(z)$. We recall that the argument developed here is valid for any graph, whether it be a regular lattice or a random network,, provided that we know at least implicitly $H_0(z)$. Of course, the practical difficulty is that in most cases we do not know $H_0(z)$, not even implicitly. In particular, no results are typically available for clustered networks. However, Eq.~\eqref{eq:H_0} is exact for the strongly clustered STC graphs.

\section{Critical exponents}
\label{sec:critical_exponents}

In this section, we compute the percolation critical exponents $\beta$, $\gamma$, and $\sigma$ for percolation in strongly clustered STC random graphs having a power-law tail with exponent $\widetilde{\gamma}_d$. We stress again the strategy we are adopting. First, we solve for the critical exponents of extended-range percolation for $R=2$ in the backbone $\mathcal{G}_0$ with power-law exponent $\gamma_d$. Hence the expressions we obtain are in terms of $\gamma_d$. Then, we connect these results to standard site percolation in $\mathcal{G}_1$, having power-law exponent $\widetilde{\gamma}_d$, via the relation $\gamma_d=\widetilde{\gamma}_d+1$.

\subsection{The exponent $\beta$}
We begin computing the critical exponents $\beta$. The key ingredient for this scope is the behavior of Eqs.~\eqref{eq:H_1},\eqref{eq:H_2} for $|t|=|\phi-\phi_c| \ll 1$ at $z=1$. Setting $m_1=1-H_1(1)$ and $m_2=1-H_2(1)$, we get from Eq.~\eqref{eq:H_1},\eqref{eq:H_2} as
\begin{align}
\label{eq:m1}
m_1 &= 1-g_1\big(1-(\phi m_1 + (1-\phi)m_2)\big),\\
\label{eq:m2}
m_2 &= 1-g_1\big(1-\phi m_1\big).
\end{align}
These equations correspond to Eqs~(10),(11) in the main with $u_1=1-m_1$ and $u_2=1-m_2$. As already discussed in the main text, these equation have a trivial solution $m_1=m_2=0$ corresponding to a non-percolating phase, and the percolation transition takes place at the threshold
\begin{equation}
\label{eq:threshold}
 1-[b \phi_c + (1-\phi_c)\phi_c b] \implies \phi_c = \frac{1+b - \sqrt{(1+b)^2-4}}{2b},
\end{equation}
where $b=\langle k(k-1) \rangle / \langle k \rangle$ is the networks branching factor. From Eq.~\eqref{eq:H_0}, we can expand using \eqref{eq:asymptotic_g0}
\begin{equation}
 P^{\infty}/\phi = 1-H_0(1) \simeq \langle k \rangle\left[\phi m_1 +(1-\phi)m_2 \right],
 \label{eq:order_asymptotic}
\end{equation}
where we neglected the singular terms with exponent $\gamma_d-1$ as they are subleading for the case $\gamma_d>2$ considered here. The critical exponent $\beta$ is determined by the behaviour of $m_1$ and $m_2$ for $t \ll 1$.

From Eq.~\eqref{eq:m1} and \eqref{eq:m2}, using Eq.~\eqref{eq:asymptotic_g1} we get
\begin{align}
\label{eq:asymptotic_m1}
  m_1 & \simeq b [\phi m_1 + (1-\phi) m_2] - \frac{1}{2}d [\phi m_1 + (1-\phi)m_2]^2 - c_2[\phi m_1 + (1-\phi)m_2]^{\gamma_d-2},\\
  \label{eq:asymptotic_m2}
  m_2 & \simeq b \phi m_1 - \frac{1}{2}\phi^2 d m_1^2 - c_2\phi^{\gamma_d-2}m_1^{\gamma_d-2}.
\end{align}
Now we need to study the solution of these equations for $\phi \simeq \phi_c$, \textit{i.e.}, $ t \ll 1$. Such a solution depends on the value of $\gamma_d$. In particular, as will be clarified below, there is a difference between the cases $\gamma_d>3$ and $2< \gamma_d <3$. Since $m_2$ depends only on $m_1$, we can substitute the second equation into the first, and keep only the lowest orders in $m_1$. In the first equation, the term $m_1^{(\gamma_d-2)^2}$ is subleading in the for $\gamma_d>3$ and dominant instead for $2<\gamma_d<3$. For $\gamma_d>3$, for which $b$ is finite and $\phi_c>0$, we get at the leading orders
 \begin{equation}
  m_1 \simeq [b \phi + (1-\phi)\phi b] m_1 + \kappa_2 \phi^2 m_1^2 + \kappa_3 \phi^{\gamma_d-2} m_1^{\gamma_d-2}, \end{equation}
where $\kappa_2$ and $\kappa_3$ are constants. In the following, as there will be many different constant and we are not interested in their explicit values, we adopt the notation $\kappa_i$ to denote them. Setting $\phi=\phi_c+t$ and using the equation of the critical threshold Eq.~\eqref{eq:threshold} we finally get
\begin{equation}
 \kappa_1 t m_1 + \kappa_2 m_1^2 + \kappa_3 m_1^{\gamma_d-2}\simeq 0.
\end{equation}
Note that $m_1=0$ is always a solution of these equations. This corresponds to $m_1=m_2=0$ and $u_1=u_2=1$, \textit{i.e.} $P^{\infty}=0$. A nontrivial solution appears only for $t>0$. Using Eq.~\eqref{eq:asymptotic_m2} we then get
\begin{equation}
 m_1 \sim m_2 \sim \begin{cases}
             t , &\gamma_d > 4\\
             t^{1/(\gamma_d-3)} &3 < \gamma_d < 4,
            \end{cases}
            \label{eq:m1_m2_gamma>3}
\end{equation}
from which we can conclude, using Eq.~\eqref{eq:order_asymptotic}, that
\begin{equation}
 \beta= \begin{cases}
                1, &\gamma_d > 4,\\
                \frac{1}{\gamma_d-3}, &3< \gamma_d < 4.
               \end{cases}
\end{equation}

For the more interesting case $2<\gamma_d<3$, when the branching factor is infinite and the threshold vanishes, we have $\phi=t$ and at lowest order
\begin{equation}
 m_1 \simeq \kappa_4 t^{(\gamma_d -2)^2} m_1^{(\gamma_d-2)^2},
\end{equation}
from which we get, using again Eq.~\eqref{eq:asymptotic_m2} with $\phi$ replaced by $t$ and keeping only the lowest order,
\begin{align}
\label{eq:m1_gamma<3}
 m_1 &\sim t^{\frac{(\gamma_d-2)^2}{1-(\gamma_d-2)^2}},\\
 \label{eq:m2_gamma<3}
 m_2 &\sim t^{\frac{\gamma_d-2}{1-(\gamma_d-2)^2}}.
\end{align}
Note that $m_1 \gg m_2$, but in all the expressions it always appear with a multiplicative factor $\phi=t$, while $m_2$ is multiplied by $(1-\phi)\sim 1$. Hence $tm_1 \ll m_2$. With this in mind, using Eq.~\eqref{eq:order_asymptotic} with $\phi=t$ we get
\begin{equation}
 P^{\infty}/t \simeq \langle k \rangle m_2 \simeq \kappa_5 t^{\frac{\gamma_d-2}{1-(\gamma_d-2)^2}},
\end{equation}
from which it follows that\footnote{In Ref.~\cite{cirigliano2023extended} the exponent associated to $P^{\infty}$ rather than the one associated to $P^{\infty}/\phi$ has been computed. This is the reason of the additional factor $1$ appearing in Eq.(9) of \cite{cirigliano2023extended}.}
\begin{equation}
 \beta = \frac{\gamma_d-2}{1-(\gamma_d-2)^2}.
\end{equation}
This exponent, already found in~\cite{cirigliano2023extended} is different from the value for $R=1$, $\beta=1/(3-\gamma_d)$ \cite{cirigliano2024scaling}, suggesting that extended-range percolation on strongly heterogeneous networks present $R$-dependent critical features and no longer belongs to the universality class of standard percolation.

\subsection{The exponent $\gamma$}
Now we turn our attention to $\langle s \rangle$, defined in Eq.~\eqref{s}, whose singular behavior is given in Eq.~\eqref{eq:scaling_exponent_3}. We take the derivatives with respect to $z$ of Eqs.~\eqref{eq:H_0}, \eqref{eq:H_1} and \eqref{eq:H_2}, evaluating them in $z=1$. Using $g_0'(x) = \langle k \rangle g_1(x)$ and the expression for $m_1$ in Eq.~\eqref{eq:m1}, from Eq.~\eqref{s} we get
\begin{align}
\langle s \rangle &= 1 + \frac{\phi \langle k \rangle u_1 \big[\phi w_1+(1-\phi)w_2 \big]}{g_0(\phi u_1 +(1-\phi)u_2)},\\
w_1 &= 1-m_1+g_1'\big(1-(\phi m_1+(1-\phi)m_2) \big) \big[\phi w_1+(1-\phi)w_2 \big],\\
w_2 &= g_1'\big(1-\phi m_1 \big)\phi w_1,
\end{align}
where we set $w_1 = H'_1(1)$ and $w_2 = H'_2(1)$. Note that the equations for $w_1$ and $w_2$ are linear equations. They can be easily solved to get
\begin{equation}
 \langle s \rangle - 1 = \frac{\phi \langle k \rangle (1-m_1)^2  \big[1 +(1-\phi)g_1'\big(1-\phi m_1 \big) \big]}{g_0(1-(\phi m_1+(1-\phi)m_2))\Theta(\phi)},
\label{eq:mean_cluster_size}
\end{equation}
where $m_1$ and $m_2$ are the solution of Eqs.~\eqref{eq:m1} and~\eqref{eq:m2}, respectively, and we defined
\begin{equation}
 \Theta(\phi) = 1-\phi g_1'\big(1-(\phi m_1+(1-\phi)m_2)\big)\big[ 1+(1-\phi)g_{1}'(1-\phi m_1)\big].
 \label{eq:Lambda_phi}
\end{equation}
The explicit dependence on $m_1$ and $m_2$ has been omitted for the sake of simplicity. Note that $\Theta(\phi_c)=0$, because of the criticality condition. Since the other term in the denominator is finite for any $\phi$, we can expand $\Theta(\phi)$ for $\phi=\phi_c+t$ and $t$ small to get the singular behaviour of the mean cluster size~\cite{cirigliano2024scaling}. First of all, note that if $t<0$, $m_1=m_2=0$ and $\Theta(\phi)=1-\phi b[1+(1-\phi)b]$, from which we simply get
\begin{equation}
\Theta(\phi) \simeq \Theta(\phi_c)+\Theta'(\phi_c) t \simeq [2\phi_cb^2-b(1+b)]t,
\end{equation}
hence from Eq.~\eqref{eq:mean_cluster_size} we get that $\gamma=1$ from the $t<0$ side of the transition. This result is valid only for $\gamma_d>3$, since for $2<\gamma_d<3$ the threshold vanishes and $t \geq 0$ always. We will compute now $\gamma$ for $t>0$. To do this, a bit more effort is required, since we need the expansion of $g_1'(1-\epsilon)$ for small $\epsilon$, given by Eq.~\eqref{eq:asymptotic_g1'}
Note that $g_1'(1-\epsilon) \to \infty $ as $\epsilon \to 0^{+}$ for $\gamma_d<3$ since the branching factor is infinite. Furthermore, the linear term is the leading one for $\gamma_d>4$, while the term proportional to $\epsilon^{\gamma_d-3}$ dominates the small $\epsilon$ behaviour for $\gamma_d<4$. With this in mind, we can now expand $\Theta(\phi)$. If $\gamma_d>4$, we can expand in Eq.~\eqref{eq:Lambda_phi} and get
\begin{align}
 \nonumber
  \Theta(\phi) &\simeq 1-\phi b [1+(1-\phi)b]+\kappa_6 m_1 + \kappa_7 m_2 \\
  &\simeq \Theta(\phi_c)t + \kappa_6 m_1 + \kappa_7 m_2 \sim t,
\end{align}
where in the last asymptotics we have used Eq.~\eqref{eq:m1_m2_gamma>3} and $\kappa_6$ and $\kappa_7$ are constants. Again, from Eq.~\eqref{eq:mean_cluster_size} we get $\gamma=1$. For $3<\gamma_d<4$, we must consider the term in $\epsilon^{\gamma_d-3}$ rather than the linear one in Eq.~\eqref{eq:asymptotic_g1'}. After a bit of algebra, we get
\begin{equation}
\Theta(\phi) \simeq \Theta(\phi_c)t + \kappa_8 [\phi_cm_1+(1-\phi_c)m_2]^{\gamma_d-3} + \kappa_9 m_1^{\gamma_d-3},
\end{equation}
where $\kappa_8$ and $\kappa_9$ are constants. Using again Eq.~\eqref{eq:m1_m2_gamma>3}, we finally get $\Theta(\phi) \sim t$, from which we conclude that $\gamma=1$.

Finally, we turn to the case $2<\gamma_d<3$. As anticipated, in this regime the branching factor is infinite and no $t<0$ phase exists. The expansion in Eq.~\eqref{eq:asymptotic_g1'} diverges as $\epsilon \to 0$. However, we can expand $\Theta(\phi)$ for $\phi=t \simeq 0$ to get at leading orders
\begin{equation}
\Theta(\phi) \simeq 1 + c_1 t m_1^{\gamma_d-3} + \kappa_2 t (tm_1)^{\gamma_d-3} + \kappa_3 t (tm_1 m_2)^{\gamma_d-3}.
\end{equation}
For $t \to 0$, the first three terms tend to $1$, because the presence of the multiplicative factor $t$ cures the divergence of the terms with the negative exponent $\gamma_d-3$. The last term must be considered carefully. Using Eq.~\eqref{eq:m1_gamma<3} and \eqref{eq:m2_gamma<3}, we have
\begin{equation}
 t (tm_1 m_2)^{\gamma_d-3} \sim t^{1+\frac{(\gamma_d-1)(\gamma_d-3)}{1-(\gamma_d-2)^2}} \sim 1,
\end{equation}
from which it follows that $\Theta(\phi) \to \text{const.}$ for $t \to 0$. No divergence is present in the denominator of Eq.~\eqref{eq:mean_cluster_size}. The singular behaviour in $t$ must be extracted then from the numerator. Using Eq.~\eqref{eq:asymptotic_g1'} we get
\begin{equation}
 \langle s \rangle -1 \simeq c_1 t + \kappa_2 t^{1+\frac{\gamma_d-3}{1-(\gamma_d-2)^2}} \sim t^{-\left[\frac{(\gamma_d-3)(\gamma_d-2)}{1-(\gamma_d-2)^2}\right]},
\end{equation}
from which we finally get
\begin{equation}
\label{eq:gamma2}
 \gamma=-\frac{(3-\gamma_d)(\gamma_d-2)}{1-(\gamma_d-2)^2}.
\end{equation}
Similarly to the case $R=1$, at the percolation threshold $\phi_c=0$ the susceptibility does not diverge, since $\gamma<0$. However, the critical exponent $\gamma \neq -1$, as it happens for $R=1$, but takes instead, surprisingly, $\gamma_d$-dependent values.

\subsection{The exponent $\sigma$}
\begin{figure}
\includegraphics[width=0.95\textwidth]{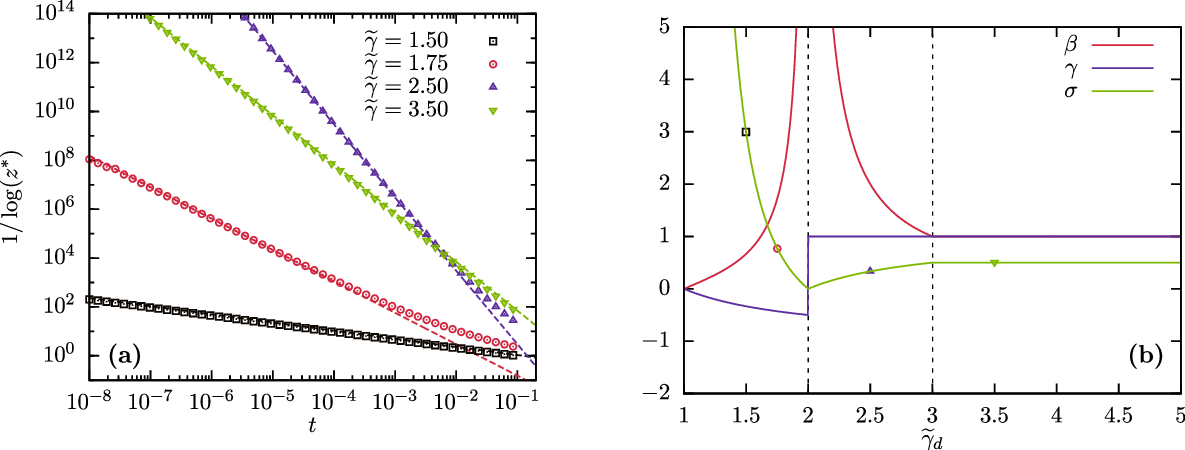}
\caption{(a) Results for $1/\log(z^*)$ for small $t=\phi - \phi_c$ from the numerical solution of Eq.~\eqref{eq:inversion} (symbols), and comparison with the predictions for the exponent $\sigma$ in Eq.~\eqref{eq:sigma} (dashed lines). (b) Plot of the critical exponents $\beta$, $\gamma$, and $\sigma$ for STC random graphs with power-law exponent $\widetilde{\gamma}_d$, from Eq.~\eqref{eq:beta_STC}-\eqref{eq:sigma_STC}. The symbols indicate the values of $\sigma$ relative to panel (a).}
\label{fig.1}
\end{figure}

We want to find the position of $z^*$, the closest singularity to the origin of $H_0(z)$. To do so, we look at the closest singularity of $H_1(z)$. The reason is related to the singular behavior of $g_0(z)$ and $g_1(z)$: since $H_0(z)$ is defined via $g_0$, and for $2<\gamma_d$ the singular behavior of $g_0$ has an exponent larger then $1$, at leading order the singular behaviour of $H_0$ is given by the linear term of $g_0$, and it is the same as the singular behaviour of $H_1$. Details can be found in~\cite{cirigliano2024scaling}. To study the singularity of $H_1$, we can write, from Eq.~\eqref{eq:H_1} and \eqref{eq:H_2}, solving for $z=H_1^{-1}(\zeta)$, the equation
\begin{equation}
 z=H_1^{-1}(\zeta)=\psi(\zeta)=\frac{\zeta}{\mathcal{G}_1\big(\zeta,\mathcal{G}_1(\zeta,1) \big)},
\end{equation}
where $\mathcal{G}_1(x,y)=g_1\big(\phi x + (1-\phi)y \big)$. Note that $
 (1-\phi)\partial_x \mathcal{G}_1(x,y)=\phi \partial_y \mathcal{G}_1(x,y)$, hence
\begin{equation}
\frac{d}{ d \zeta }\mathcal{G}_1\big(x(\zeta),y(\zeta)\big)= \frac{\partial_x \mathcal{G}_1\big(x(\zeta),y(\zeta) \big)}{\phi}\left[\phi x'(\zeta)+(1-\phi)y'(\zeta) \right].
\end{equation}
The inversion of $\psi$ works until we reach its first stationary point. This defines the characteristic equation $\psi'(\zeta^*)=0$, which writes then as
\begin{equation}
\label{eq:inversion}
 \left[\mathcal{G}_1\big(\zeta,\mathcal{G}_1(\zeta,1) \big)-\zeta \frac{d}{d \zeta}\mathcal{G}_1\big(\zeta,\mathcal{G}_1(\zeta,1) \big)\right]_{\zeta=\zeta^*}=0,
\end{equation}
where
\begin{widetext}
\begin{align*}
 \frac{d}{d \zeta}\mathcal{G}_1\big(\zeta^*,\mathcal{G}_1(\zeta^*,1) \big)&=\phi g_1'\big(\phi \zeta^*+(1-\phi)g_1(\phi \zeta^* + 1-\phi)\big)\left[1+(1-\phi)g_1'(\phi \zeta^* +1-\phi) \right].
\end{align*}
\end{widetext}
The solution of Eq.~\eqref{eq:inversion} gives us the point $\zeta^{*}$ at which we can no longer invert the function $\psi$ to get $H_1$. This is reflected in a singularity at $z^*=\psi(\zeta^*)$. This singularity is such that $z^* \to 1$ as $t \to 0$, and we can write $z^{*} \sim 1 + \theta(t)$, were $\theta(t) \to 0$ as $t\to 0$.
The correlation size is given by
\[s_{\xi}=\frac{1}{\log(z^*)}\sim \frac{1}{\log(1+\theta(t))}\sim \theta(t)^{-1}. \]
In principle, one can solve Eq.~\eqref{eq:inversion} at leading orders for $t\ll1$, plug the corresponding $\zeta^*$ into $z=\psi(\zeta^*)$, and expand again at leading order to get the exponent $\sigma$. Here, we follow a simpler strategy. First, we use the scaling relation $\sigma=1/(\beta+\gamma)$ to get
\begin{equation}
\label{eq:sigma}
 \sigma= \begin{cases}
                \frac{1}{2}, &\gamma_d > 4,\\
                \frac{\gamma_d-3}{\gamma_d-2}, &3< \gamma_d < 4,\\
                \frac{1-(\gamma_d-2)^2}{(\gamma_d-2)^2},\quad&2<\gamma_d<3,
               \end{cases}
\end{equation}
Then we just solve numerically Eq.~\eqref{eq:inversion} for small $t$ and compare $1/\log(z^*)$ with our predictions in Eq.~\eqref{eq:sigma}, showing perfect agreement (see Fig.~\ref{fig.1}(a)) and proving that scaling relations hold.

\subsection{Critical exponents for the STC networks}
The last step is to map our results for extended-range percolation on the backbone $\mathcal{G}_0$ to standard site percolation on the STC graph $\mathcal{G}_1$. Setting $\gamma_d=\widetilde{\gamma}_d+1$, we have
\begin{align}
\label{eq:beta_STC}
 \beta&= \begin{cases}
                1, \quad&\widetilde{\gamma}_d > 3,\\
                \frac{1}{\widetilde{\gamma}_d-2}, \quad&2< \widetilde{\gamma}_d < 3,\\
                \frac{\widetilde{\gamma}_d-1}{1-(\widetilde{\gamma}_d-1)^2},\quad&1<\widetilde{\gamma}_d<2,
               \end{cases}\\
\label{eq:gamma_STC}
\gamma&= \begin{cases}
                1, \quad&\widetilde{\gamma}_d > 2,\\
                -\frac{(2-\widetilde{\gamma}_d)(\widetilde{\gamma}_d-1)}{1-(\widetilde{\gamma}_d-1)^2},\quad&1<\widetilde{\gamma}_d<2,
               \end{cases}\\
\label{eq:sigma_STC}
\sigma&= \begin{cases}
                \frac{1}{2}, \quad&\widetilde{\gamma}_d > 3,\\
                \frac{\widetilde{\gamma}_d-2}{\widetilde{\gamma}_d-1}, \quad&2< \widetilde{\gamma}_d < 3,\\
                \frac{1-(\widetilde{\gamma}_d-1)^2}{(\widetilde{\gamma}_d-1)^2},\quad&1<\widetilde{\gamma}_d<2,
               \end{cases}
\end{align}
which are reported in Table 1 of the main text. See also Fig.~\ref{fig.1}(b).

\appendix
\section{The generating functions of power-law distributions}
As it is needed in all the computations developed here, we report for completeness the singular behaviour for $z \to 1^+$, \textit{i.e.}, for $\epsilon=1-z\ll 1$, of the generating function $g_0(z)=\sum_{k}p_kz^k$ of power-law degree distributions $p_k \sim k^{-\gamma_d}$. Such singular behaviour is given by~\cite{cirigliano2024scaling}
\begin{align}
\label{eq:asymptotic_g0}
 g_0(1-\epsilon) \simeq 1 - \langle k \rangle \epsilon + \frac{1}{2}\langle k \rangle b \epsilon^2 + c_1 \epsilon^{\gamma_d-1} ,
\end{align}
and similarly for $g_1(z)=g_0'(z)/g'_0(1)$ and its derivative
\begin{align}
\label{eq:asymptotic_g1}
 g_1(1-\epsilon) &\simeq 1 - b\epsilon + \frac{1}{2} d \epsilon^2 + c_2 \epsilon^{\gamma_d-2},\\
 \label{eq:asymptotic_g1'}
 g_1'(1-\epsilon) &\simeq b -d \epsilon + c_3 \epsilon^{\gamma_d-3},
\end{align}
where $\langle k \rangle$ is the average degree, $b=\langle k(k-1)/\langle k\rangle$ is the network branching factor if $\gamma_d>3$, and a negative constant if $2<\gamma_d<3$, $d=\langle k(k-1)/\langle k\rangle$ if $\gamma_d>4$, and a constant whose value and sign depend on $\gamma_d$ otherwise, and $c_1$, $c_2$, and $c_3$ are constant which depend on $\gamma_d$ and on the details of the degree distributions.

\bibliography{2_references}